\documentclass[11pt,a4paper]{article}
\usepackage[latin1]{inputenc}
\usepackage[T1]{fontenc}
\usepackage{amssymb}
\usepackage{amsmath}
\usepackage{amsfonts}
\usepackage{amsthm,color}
\usepackage{hyperref}
\usepackage[margin=1in]{geometry} 

\makeatletter \@addtoreset{equation}{section}

\renewenvironment{thebibliography}[1]{%
\begin{oldthebibliography}{#1}%
\setlength{\baselineskip}{.9em}
\linespread{1}
\small
\setlength{\parskip}{0.3ex}%
\setlength{\itemsep}{.5em}%
}%
{%
\end{oldthebibliography}%
}

\def \benumlab#1{\begin{enumerate}[label={\rm \bf{(#1{\arabic{*}})}}, ref={\rm #1{\arabic{*}}}]}
\def \enumlab{\end{enumerate}}
\def \benumlabi#1{\begin{enumerate}[label={\rm {(#1\roman{*})}}, ref={\rm{(#1\roman{*})}}]}

\def\Lb{{\mathbf L}}

\def \E{\mathbb{E}}
\def \F{\mathbb{F}}

\def \R{\mathbb{R}}

\def\P{\mathbb{P}}
\def\Q{\mathbb{Q}}

\def\Bc{{\cal B}}

\def\Ec{{\cal E}}
\def\Fc{{\cal F}}

\def\Lc{{\cal L}}

\def\Rc{{\cal R}}

\def\ep{\hbox{ }\hfill$\Box$}
\def\reff#1{{\rm(\ref{#1})}}
\def\be{\begin{eqnarray}}
\def\ee{\end{eqnarray}}
\def\bal{\begin{aligned}}
\def\eal{\end{aligned}}
\def\beq{\begin{equation}}
\def\eeq{\end{equation}}
\def\beqq{\begin{equation*}}
\def\eeqq{\end{equation*}}
\def\b*{\begin{eqnarray*}}
\def\e*{\end{eqnarray*}}

\def\={\;=\;}

\def\.{\;.}

\def\eps{\varepsilon}

\def\vp{\varphi}
\def\1{\mathbf{1}}

\theoremstyle{plain}
\newtheorem{theorem}{Theorem}[section]
\newtheorem{proposition}[theorem]{Proposition}
\newtheorem{corollary}[theorem]{Corollary}
\newtheorem{assumption}[theorem]{Assumption}

\newtheorem{lemma}[theorem]{Lemma}
\theoremstyle{definition}
\newtheorem{remark}[theorem]{Remark}
\newtheorem{example}[theorem]{Example}

\def\vc{\vartheta}

 \def\L{{\mathbf L}}
 \def\No#1#2{ \|#2 \|_{\L_{#1}(\Q)}}

\def\bru#1{{\color{blue}{#1}}}
\begin{document}

\title{Simple Bounds for Utility Maximization\\ with Small Transaction Costs\footnote{The authors are grateful to Costas Kardaras and Mihai Sirbu for fruitful discussions.}}
\author{Bruno Bouchard\thanks{Universit\'e Paris-Dauphine, PSL Research University, CNRS, UMR [7534], CEREMADE, email \texttt{bouchard@ceremade.dauphine.fr}.}
\and
Johannes Muhle-Karbe\thanks{Imperial College London, Department of Mathematics, email \texttt{j.muhle-karbe@imperial.ac.uk}.}
}
\date{\today}
\maketitle

\begin{abstract}
Using elementary arguments, we show how to derive $\L_{p}$-error bounds for the approximation of frictionless wealth process in markets with proportional transaction costs. For utilities with bounded risk aversion, these estimates yield lower bounds for the frictional value function, which  pave the way for its asymptotic analysis using stability results for viscosity solutions. Using tools from Malliavin calculus, we also derive simple sufficient conditions for the regularity of frictionless optimal trading strategies, the second main ingredient for the asymptotic analysis of small transaction costs. 
\end{abstract}

\vspace{2mm}


\vspace{2mm}

\noindent {\bf JEL Clasification:}  G11, C61
\section{Introduction}

Transaction costs, such as bid-ask spreads, are a salient feature of even the most liquid financial markets. Their presence substantially complicates financial decision making by introducing a nontrivial tradeoff between the gains and costs of trading. Indeed, with transaction costs, the position in each asset is no longer a control variable that can be specified freely. Instead it becomes an additional state variable that can only be adjusted gradually. Consequently, models with transaction costs are notoriously intractable and rarely admit explicit solutions even in the simplest concrete settings~\cite{constantinides.magill.76,davis.norman.90,shreve.soner.94}. 

As a way out, one can view models with transaction costs as perturbations of their frictionless counterparts, and study their asymptotics around these more tractable benchmarks. This asymptotic point of view was first used to obtain closed-form approximations in simple concrete models, cf., e.g., \cite{shreve.soner.94,whalley.wilmott.97,janecek.shreve.04,bichuch.shreve.13,bichuch.14,gerhold.al.14}. More recently, these results have been extended to increasingly more general settings \cite{soner.touzi.13,possamai.al.15, kallsen.muhlekarbe.17,martin.14,kallsen.li.13,bouchard.al.16,melnyk.seifried.17}.

Rigorous convergence proofs for such small-cost asymptotics are typically either based on stability results for viscosity solutions or on convex duality.\footnote{A related approach, building on \cite{fukasawa.14,gobet.landon.14}, studies tracking errors using weak-convergence techniques \cite{cai.al.17}.} For the first approach, pioneered by \cite{soner.touzi.13}, the starting point for the analysis is that the difference between the value functions with and without transaction costs indeed admits an expansion of a certain asymptotic order. In simple models where explicit calculations are possible, this assumption has been verified in \cite{soner.touzi.13,possamai.al.15,bouchard.al.16} by constructing explicit subsolutions of the respective frictional dynamic programming equations.  In a related model with quadratic costs \cite{moreau.al.17}, the corresponding bound is established using a classical verification argument under very strong additional regularity conditions that, however, rule out standard portfolio choice models such as \cite{kim.omberg.96}.

In the papers based on convex duality \cite{kallsen.li.13,herdegen.muhlekarbe.17}, a lower bound is derived by considering a specific almost optimal control. This is in turn complemented by constructing a corresponding dual element, for which the bound is tight at the leading asymptotic order for small transaction costs. This approach again requires strong regularity conditions, in particular on the frictionless optimizer. These are generally not easy to verify and only satisfied for sufficiently short time horizons in the model of \cite{kim.omberg.96}, for example. 

In the present paper, we show how bounds for utility maximization problems with transaction costs can be derived using arguments that are simple and elementary, but nevertheless apply to the model of \cite{kim.omberg.96}, for example. The 
expected asymptotic order formally arises as the optimal trade-off between displacement from the frictionless optimizer and the cost of tracking it, cf.~\cite{Rogers04} and \cite[Remark 4]{janecek.shreve.04}. We show that this idea can be exploited to obtain rigorous $\L_{p}$ estimates for the corresponding tracking errors. Combined with a simple trick from \cite[Proof of Theorem~3.1]{BEM12}, this directly leads to the desired bounds for utility maximization problems\footnote{In order not to drown the main ideas in generality, we only consider utility functions defined on the real line in the present paper. Power-like utility functions (with bounded \emph{relative} risk aversion) defined on the positive halfline can be treated with quite similar arguments, at the cost of somewhat more involved estimates.}. Once the correct asymptotic order is identified using this bound, the corresponding leading-order term can in turn be determined using the viscosity approach of \cite{soner.touzi.13,possamai.al.15,bouchard.al.16}. 

Our arguments require mild integrability conditions, some of which are expressed in terms of the frictionless optimizer. However, using techniques from Malliavin calculus, we show that in complete Markovian\footnote{A similar approach can be used in a non-Markovian setting but we do not consider this general case to keep the arguments simple.} markets these can be easily verified in terms of the primitives of the model. 

The remainder of the article is organized as follows. Our model with proportional transaction costs is introduced in Section~\ref{s:model}. Subsequently, in Section~\ref{s:bounds}, we derive simple pathwise bounds for the transaction costs accumulated when tracking frictionless target strategies by the solutions of simple Skorohod problems. Under mild integrability conditions, these in turn lead to $\L_{p}$-error bounds for the approximation of frictionless wealth process in markets with proportional transaction costs. In Section~\ref{s:utility}, we use these results to derive upper and lower bounds for utility maximization problems with transaction costs. Using tools from Malliavin calculus, simple sufficient conditions for the validity of these results are provided in Section~\ref{s:malliavin}. Finally, in Section~\ref{sec: TC on amount}, we discuss how to extend our approach to transaction costs proportional to monetary amounts rather than numbers of shares traded.

\section{Model}\label{s:model}

Let $(\Omega,\Fc,\P,\F=(\Fc_{t})_{t \in [0,T]})$ be a filtered probability space satisfying the usual conditions. We consider a financial market with $1+d$ assets. The first is safe, with price normalized to one. The other $d$ assets are risky, with prices modeled by a $\R^{d}$-valued continuous semimartingale $S$. 

Without transaction costs, trading strategies are described by  $\mathbb{R}^d$-valued predictable $S$-integrable processes $\theta$. Here, $\theta^i_t$ denotes the number of shares of risky asset $i$ held at time $t$. Accordingly, the frictionless wealth process corresponding to a strategy $\theta$ and the fixed initial endowment $X_0 \in \R$ is
$$
X^\theta_t :=X_{0}+\int_{0}^t  \theta_{s}^{\top} dS_{s}, \quad t \in [0,T].
$$ 
Now suppose as in \cite{janecek.shreve.10,bichuch.shreve.13,martin.14} that trades incur costs proportional to the number of units transacted.\footnote{Transaction costs proportional to the dollar amounts traded as in \cite{{constantinides.magill.76},davis.norman.90,shreve.soner.94} can be treated along the same lines, leading to somewhat more involved integrability conditions; cf.~Section \ref{sec: TC on amount} below for more details.} Then, trading strategies $\vartheta$ necessarily have to be of finite variation and the frictional wealth process corresponding to a $\R^{d}$-valued predictable, c\`adl\`ag, finite-variation process with initial value $\vartheta_{0-}=0$ is
\begin{equation}\label{eq:wealtheps}
X^{\vartheta,\eps}_t :=X_{0}+\int_{0}^{t}  \vartheta_{s}^{\top}dS_{s}- \eps\int_{0}^{t}d|\vartheta|_{s}- \1_{\{T\}}\eps |\vartheta_{T}|.
\end{equation}
Here, $\varepsilon>0$ is the proportional transaction cost  and $|\vartheta|:=\sum_{i=1}^{d} |\vartheta^{i}|$, where $|\vartheta^{i}|_t$ is the total variation of the position $(\vartheta^{i}_s)_{s\in [0,t]}$ in the risky asset $i$ on $[0,t]$. Buying or selling a number $d\vartheta^{i}_{t}$ of units risky asset $i$ at time $t$ induces the transaction costs $\varepsilon d|\vartheta^{i}|_{t}$ in \eqref{eq:wealtheps}. Likewise, the term $\1_{\{T\}}\eps  |\vartheta_{T}|$ describes the transaction cost paid when  liquidating the risky asset positions at the terminal time $T$. 

In this paper, we show that elementary arguments allow to bound the error made by approximating the frictionless wealth process $X^\theta$ by a frictional wealth process $X^{\vartheta,\varepsilon}$ when $\vartheta$ is defined as the solution of a Skorokhod problem\footnote{Here, we use the canonical decomposition $\vartheta^{i}=\vartheta^{i+}-\vartheta^{i-}$ in which $\vartheta^{i\pm}$ is c\`adl\`ag and non-decreasing.} for some $\delta>0$:
\begin{equation}\label{eq: skoroh}
\left\{\begin{array}{c}
\theta-\vartheta\in [-\delta,\delta]^{d} \mbox{ on } [0,T],\\
\sum_{i=1}^{d}\left(\int_{0}^{T}\1_{\{\theta^{i}_{t}-\vartheta^{i}_{t}=\delta \}}d\vartheta^{i +}_{t}+\int_{0}^{T}\1_{\{\theta^{i}_{t}-\vartheta^{i}_{t}=-\delta\} }d\vartheta^{i-}_{t}\right)=0.
\end{array}\right.
\end{equation} 
This means that the position $\vartheta^i_t$ in each risky asset is held constant as long as it differs from the frictionless target allocation $\theta^i_t$ by no more than $\delta$. Once this threshold is reached, just enough trading is performed to maintain a deviation smaller than or equal to $\delta$.

 \section{Bounds for the Skorokhod Problem and Tracking Error}\label{s:bounds}

Throughout, we assume that the frictionless target strategy $\theta$ is a continuous semimartingale,\footnote{In Section~\ref{s:malliavin}, we use tools from Malliavin calculus to provide sufficient conditions for this assumption in terms of the primitives of the model.} and use the shorthand notation $\langle \theta\rangle:=\sum_{i=1}^{d}\langle \theta^{i}\rangle$.  The first step to derive our tracking-error estimates are the following elementary pathwise bounds for the transaction costs accumulated by the frictional tracking strategy \eqref{eq: skoroh}. Their derivation is based on a  simple application of It\^o's formula reminiscent of \cite[Remark 4]{janecek.shreve.04}. 

Here and henceforth, $\Bc_{1}$ denotes the collection of predictable processes with values in $[-1,1]^{d}$.

\begin{lemma}\label{lem: estimate L1} 
Fix $\delta\in (0,1)$ and let $\vartheta$ be the solution of the Skorohod problem \eqref{eq: skoroh}. Then, there exists $\xi \in \Bc_{1}$ such that  
\begin{align}\label{eq: bound L in terms of processes}
|\vartheta|_{t}\le \Rc_{\delta}(\xi)_{t}:=2d\delta+ \int_{0}^{t}   \xi_{s}^{\top}d\theta_{s} +\frac{1}{2\delta }  \langle \theta\rangle_{t}, \quad \mbox{ for all $t\in [0,T]$.} 
 \end{align}
\end{lemma}

 \proof Since we are working on the rectangle $[-\delta,\delta]^{d}$, we can consider each component separately and then sum the respective bounds to obtain \eqref{eq: bound L in terms of processes}. Without loss of generality, we therefore suppose $d=1$ and consider a smooth function with bounded derivatives $\vp$ such that 
$-\vp'(-1)=\vp'(1)=1$, and $|\vp|\vee |\vp'|\vee |\vp{''}|\le 1$ on $[-1,1]$. Define the adapted, $[-1,1]$-valued process $Z:=(\theta-\vartheta)/\delta$. It\^{o}'s formula, the dynamics \eqref{eq: skoroh}, and $-\vp'(-1)=\vp'(1)=1$ give
\begin{align}
 \vp(Z_{t})=&\vp(Z_{0})+\frac1\delta\left(\int_{0}^{t}   \vp'( Z_{s})d(\theta-\vartheta)_{s}
+\frac1{2\delta } \int_{0}^{t}  \vp{''}(Z_{s})d\langle \theta\rangle_{s}\right)\nonumber\\
=&\vp(Z_{0})+\frac1\delta\left(\int_{0}^{t}   \vp'( Z_{s})d\theta_{s}-|\vartheta|_{t}
+\frac1{2\delta } \int_{0}^{t}  \vp{''}(Z_{s})d\langle \theta\rangle_{s}\right).\label{eq : proof lemma estimate L}
 \end{align}
Since $\vp$ and its first and second-order derivatives are bounded by $1$, this yields \eqref{eq: bound L in terms of processes} for $d=1$. For several risky assets, the corresponding estimates follow by summing these bounds over all $d$ components.
 \ep\\

Now, fix a probability measure $\Q$ equivalent to the physical probability $\P$. In applications to utility maximization problems, this will be the frictionless dual martingale measure that minimizes the dual problem for the original optimization; cf.~Section~\ref{s:utility}.  Another natural choice is $\P$ itself. The pathwise estimates from Lemma~\ref{lem: estimate L1} yield $\L_{p}(\Q)$-estimates under the following mild integrability conditions on the frictional target strategy:
 
\begin{assumption}\label{ass: Lp integrability} 
For some $p \ge 1$, there exists a constant $C_{\ref{eq: Lp integrability}}(p)>0$ such that
\begin{align}\label{eq: Lp integrability}
\No{p}{\langle \theta\rangle_{T}} +\sup_{\xi\in \Bc_{1}}\Big|\Big|\int_{0}^{T}   \xi_{s}^{\top}d\theta_{s}\Big|\Big|_{\mathbf{L}_p(\Q)}
\le C_{\ref{eq: Lp integrability}}(p).
\end{align} 
\end{assumption}

\begin{remark}
Suppose the frictionless target strategy is an It\^o process with dynamics 
$$d\theta_t=\mu^\theta_t dt+\sigma^\theta_t dW_t,$$
for a $\Q$-Brownian motion $W$. Then the inequalities of Minkowski, Jensen, and Burkholder-Davis-Gundy show that the bound \eqref{eq: Lp integrability} is satisfied if
\begin{equation}\label{eq:suff1}
\int_0^T \mathbb{E}^{\Q}\left[|\mu_t^\theta|^p+|\sigma_t^\theta|^p\right]dt <\infty.
\end{equation}
\end{remark} 

Under Assumption~\ref{ass: Lp integrability} , the following $\L_{p}(\Q)$-estimates are a direct consequence of Lemma~\ref{lem: estimate L1}:

\begin{corollary}\label{cor: estimate L} 
Fix $\delta\in (0,1)$ and let $\vartheta$ be the solution of the Skorohod problem \eqref{eq: skoroh}. For $p\ge1$ as in Assumption \ref{ass: Lp integrability}, there exists a constant $C_{\ref{eq: Lp bound on L}}(p)>0$ such that 
\begin{align}\label{eq: Lp bound on L}
\No{p}{\hspace{0.4mm} |\vc|_{t} \hspace{0.4mm}} &\le C_{\ref{eq: Lp bound on L}}(p)\left(  1+   \frac1{\delta} \right), \quad \mbox{ for all  $t\in [0,T]$.}
 \end{align}
\end{corollary}
 
 \begin{remark}\label{rem: Lp error is sharp}  Assume that $\theta$ is a $\Q$-Brownian motion and choose $\vp(z)=z^{2}/2$ for $z\in [-1,1]$ in the proof of Lemma \ref{lem: estimate L1}. Then, taking the expectation in \eqref{eq : proof lemma estimate L} leads to 
 $$
\E^{\Q}\left[|\vartheta|_{t}\right]=\E^{\Q}\left[\delta(\vp(Z_{0})-\vp(Z_{t}))+\frac{d}{2\delta}{t}\right]\ge -\frac12 +\frac{d}{2\delta}{t}
 $$
 for $\delta \in (0,1)$. This shows that the estimate \eqref{eq: Lp bound on L} in terms of $1/\delta$ can not be improved in general, up to constants.  
 \end{remark} 

 As a corollary, we now deduce the $\L_{p}(\Q)$-error made when approximating $X^\theta$ by $X^{\vartheta,\varepsilon}$, where $\vartheta$ is defined as in Lemma \ref{lem: estimate L1}. This requires the following additional integrability assumption on the price process $S$:

\begin{assumption}\label{ass: Lp integrability dS/S} Given $p\ge 1$  as in Assumption \ref{ass: Lp integrability}, there is a constant $C_{\ref{eq: Lp bound dS/S}}(p)>0$ such that 
\begin{align}\label{eq: Lp bound dS/S}
\sup_{\xi\in \Bc_{1}} \Big|\Big|\int_{0}^{T}   \xi_{t}^{\top}dS_{t}\Big|\Big|_{\mathbf{L}_p(\Q)} \le C_{\ref{eq: Lp bound dS/S}}(p).
\end{align}  
\end{assumption}

\begin{remark}\label{rem: cas S ito}If $S$ is $\Q$-martingale and $p\ge 1$, then \eqref{eq: Lp bound dS/S} is equivalent to $\E^{\Q}[(\langle S\rangle_{T} )^{\frac{p}2}]<\infty$ by the  Burkholder-Davis-Gundy inequality. More generally, if the returns have It\^o dynamics 
$$dS_t=\mu^S_tdt+\sigma^S_t dW_t,$$
for a  {$\Q$-}Brownian motion $W$, then the inequalities of Minkowski, Jensen, and Burkholder-Davis-Gundy show that a sufficient condition for \eqref{eq: Lp bound dS/S} is
\begin{equation}\label{eq:suff2}
\int_0^T \mathbb{E}^{\Q}\left[|\mu_t^S|^p+|\sigma_t^S|^p\right]dt <\infty.
\end{equation}
 \end{remark} 

\begin{theorem}\label{thm: estimate V-Veps} 
Define $\vartheta$ as in Lemma \ref{lem: estimate L1} with $\delta\in (0,1)$.  Then, there exist $\xi,\xi' \in \Bc_{1}$ such that 
\begin{equation}\label{eq:Rbar}
\left|X^{\vartheta,\eps}_{t}-X^\theta_{t}\right| \le  \bar \Rc_{\delta,\varepsilon}(\xi,\xi')_{t}:=\delta \left|\int_{0}^{t} \xi_{s}^{\top} dS_{s}\right| +  2 \varepsilon \Rc_{\delta}(\xi')_{t}, \quad \mbox{ for all $t \in [0,T]$.}
\end{equation}
If moreover,  Assumptions \ref{ass: Lp integrability} and \ref{ass: Lp integrability dS/S} hold, then 
\begin{align*}
\No{p}{X^{\vartheta,\eps}_{T}-X^\theta_{T}}&\le  \delta\; C_{\ref{eq: Lp bound dS/S}}(p) +  \;\varepsilon\; 2C_{\ref{eq: Lp bound on L}}(p)\left(  1+   \frac1{\delta} \right).
\end{align*}
In particular, for $\delta=\varepsilon^{1/2}\in (0,1)$, there exists a constant  $C_{\ref{eq: LP bound on diff of wealth for delta = sqrt epsilon}}(p)>0$ such that
\begin{align}\label{eq: LP bound on diff of wealth for delta = sqrt epsilon}
\No{p}{X^{\vartheta,\eps}_{T}-X^\theta_{T}}&\le C_{\ref{eq: LP bound on diff of wealth for delta = sqrt epsilon}}(p) \;\varepsilon^{1/2}.
\end{align} 
\end{theorem}

\proof By definition of the frictionless and frictional wealth processes $X^\theta$, $X^{\vartheta,\varepsilon}$ and \eqref{eq: skoroh},  
\begin{align*}
\left|X^{\vartheta,\varepsilon}_{t}-X^\theta_{t}\right|&=\left|\int_{0}^{t} (\vartheta_{s}-\theta_{s})^{\top} dS_{s}- \varepsilon|\vartheta|_{t}- \1_{\{T\}}\varepsilon |\vartheta_{T}| \right| \le \delta \left|\int_{0}^{t} \xi_{s}^{\top} dS_{s}\right| +  2\varepsilon|\vartheta|_{t},
\end{align*}
where $\xi:=(\vartheta-\theta)/\delta \in \Bc_{1}$. The claims now follow from \eqref{eq: bound L in terms of processes}, \eqref{eq: Lp bound on L}, and Assumption \ref{ass: Lp integrability dS/S}.
\ep

\begin{remark}\label{rem: weak error on wealth} 
In the context of Remarks \ref{rem: Lp error is sharp} and \ref{rem: cas S ito}, one can be more precise:
$$
\delta \E^{\Q}\left[\int_{0}^{t} \xi^{\top}_{s} \mu_{s}^{S}ds\right]-c\varepsilon\left(1+\frac1\delta\right) \le \E^{\Q}\left[X^{\vartheta,\varepsilon}_{t}-X^\theta_{t}\right]\le   \delta \E^{\Q}\left[ \int_{0}^{t} \xi^{\top}_{s} \mu_{s}^{S}ds\right] -c'\varepsilon\left(1+\frac1\delta\right),
$$
for some constants $c,c'>0$ and $\xi:=(\vartheta-\theta)/\delta \in \Bc_{1}$. Unless  $\E^{\Q}[\int_{0}^{t} \xi^{\top}_{s} \mu_{s}^{S}ds]$ vanishes, the  two error terms are of the same order only when $\delta$ is of the order of $\varepsilon^{1/2}$. This is in particular the case if $\theta$ is just a $\Q$-Brownian motion and $\mu^{S}$ is a positive constant. 
\end{remark}

\begin{remark}\label{rem:martingale}
In applications to utility maximization, $\Q$ typically is a dual martingale measure, i.e., $S$ is a $\Q$-martingale, compare Section~\ref{s:utility}. In this case, $\delta=\varepsilon^{1/2}$ no longer yields the optimal tradeoff. In this context, we will instead use the following estimate that follows from Lemma \ref{lem: estimate L1}:
$$
\E^{\Q}[X^{\vartheta,\varepsilon}_{t}-X^\theta_{t}]=- \varepsilon \E^{\Q}[ |\vartheta|_{t}+ \1_{\{T\}}|\vartheta_{T}|]\ge - 2  \varepsilon\E^{\Q}[\Rc_{\delta}(\xi)_{t}].
$$
\end{remark}


\section{Bounds for Utility Maximization with Transaction Costs}\label{s:utility}

We now apply the bounds from Section~\ref{s:bounds} to expected utility maximization problems. We focus on utility functions with bounded risk aversion defined on the whole real line:\footnote{As mentioned in the introduction, power-like utility functions could also be considered using similar arguments. However, as usual, they have to be treated separately and lead to somewhat more involved estimates.}

\begin{assumption}\label{ass : bounded risk aversion} A \emph{utility function} is a mapping $U:\R\to \R$ that is strictly increasing, strictly concave, $C^{2}$, and has bounded absolute risk aversion:
\begin{equation}\label{eq:ARA}
0<r< -\frac{U''(x)}{U'(x)}<R<\infty, \quad \mbox{for constants $r$, $R$ and all $x \in \R$.}
\end{equation}
\end{assumption}

\begin{remark}\label{rem: inada et ae} The condition \eqref{eq:ARA} implies that the derivative of $\ln(U')$ takes values in the interval $[-R,-r]\subset (-\infty,0)$. In particular, with the convention  $U(0)=0$, 
\begin{align*}
e^{-Rx+c}\le  U'(x)\le e^{-rx+c}\; \mbox{ and }\;   -\frac1R e^{-Rx+c}\le  U(x)\le  -\frac1r e^{-rx+c},\quad x\in \R,
\end{align*}
where $c:=\ln(U'(0))$. This readily implies the Inada and reasonable asymptotic elasticity conditions required for the validity of existence and duality results like \cite[Theorem 1]{schachermayer.03}.
\end{remark}

As observed by \cite{sirbu.09}, admissible strategies for such utilities with bounded absolute risk aversion can be defined as for exponential utilities in \cite{delbaen.al.02,schachermayer.03} by requiring the wealth process $X^\theta$ of frictionless admissible strategies to be supermartingales under all absolutely continuous martingale measures with finite entropy.\footnote{Indeed, in view of \eqref{eq:ARA}, the convex dual of $U$ and the relative entropy (the convex dual of the exponential utility) can be written as a convex function of each other in this case. By Jensen's inequality, this notion of admissibility thus coincides with the set $\mathcal{H}_2$ of \cite{schachermayer.03}; in particular, it does not depend on the initial endowment.} We denote the set of all such admissible strategies by $\mathcal{A}$. Under the no-arbitrage assumption 
\begin{equation}\label{eq:NA}
\left\{ \Q  \sim \P:  \E\left[ \tfrac{d\Q}{d\P} \ln (\tfrac {d\Q}{d\P}) \right]<\infty \mbox{ and } S \mbox{ is a local $\Q$-martingale}\right\} \neq \emptyset,
\end{equation}
and using that the existence of a dual minimizer follows from~\cite[Propositions 3.1 and 3.2]{kabanov.stricker.02} under Assumption~\ref{ass : bounded risk aversion}, \cite[Theorem 1]{schachermayer.03} shows that there exists an optimizer $\widehat{\theta} \in \mathcal{A}$ of the frictionless utility maximization problem 
$$
\theta \mapsto u(\theta):=\E[U(X^\theta_T)].
$$
This optimizer is related to the solution $\widehat{\Q}$ of a corresponding dual minimization problem \cite[Equation~(7)]{schachermayer.03} by the following first-order condition \cite[Equation (12)]{schachermayer.03}:
\begin{equation}\label{eq:FOC}
\frac{U'(X^{\widehat\theta}_{T})}{\E[U'(X^{\widehat\theta}_{T})]}=\frac{d\widehat\Q}{d\P}. 
\end{equation}
Here, $\widehat{\Q}$ is an equivalent (local) martingale measure for $S$. For the utility maximization problem with transaction costs,
$$
\vartheta \mapsto u^\varepsilon(\vartheta):= \E[U(X^{\vartheta,\varepsilon}_T)],
$$
admissibility can be defined in direct analogy, by requiring frictional wealth processes to be supermartingales under any absolutely continuous martingale measure with finite entropy.\footnote{In Markovian diffusion settings, this notion of admissibility  allows to apply the arguments of \cite[Theorem~3.5]{weakdpp} to obtain the weak version of the dynamic programming principle satisfied by the corresponding value function $v^{\varepsilon}$. This in turn leads to the characterization of   $v^{\varepsilon}$ as a (possibly discontinuous) viscosity super- and subsolution of a quasi-variational parabolic differential equation that is used in the above mentioned papers.} We write $\mathcal{A}^\varepsilon$ for the set of these admissible frictional strategies and note that $\mathcal{A}^\varepsilon \subset \mathcal{A}$ by, e.g., \cite[Lemma~E.5]{herdegen.muhlekarbe.17}. Since the transaction costs are always nonnegative, the frictionless value function in turn provides a natural upper bound for its frictional counterpart:
\begin{equation}\label{eq:upper}
v := \sup_{\theta \in \mathcal{A}} u(\theta) \geq \sup_{\vartheta \in \mathcal{A}^\varepsilon} u^\varepsilon(\vartheta) =: v^\varepsilon. 
\end{equation}
In \cite{soner.touzi.13,bouchard.al.16,possamai.al.15,moreau.al.17}, stability results for viscosity solutions are used to characterize the asymptotics of the frictional value function $v^\varepsilon$ for small transaction costs $\varepsilon$ in a Markovian framework. The starting point for these analyses is the abstract assumption that the normalized difference $(v-v^\varepsilon)/\varepsilon^{2/3}$ between the frictionless and frictional value functions is locally uniformly bounded with respect to the initial time and space conditions.

We now establish such a bound by using the estimates from Section~\ref{s:bounds} to complement the lower bound \eqref{eq:upper} with an appropriate upper bound. In order to apply the results from Section~\ref{s:bounds}, we need to assume that the frictionless optimizer is a continuous semimartingale; using tools from Malliavin calculus, sufficient conditions for this assumption are derived in Section~\ref{s:malliavin} in a typical Markovian setting. In addition to the integrability conditions from Section~\ref{s:bounds}, we also need   some (arbitrarily small) exponential moments to be finite:

\begin{assumption}\label{ass: U=gene}  
There exist $\iota>0$ and $C_{\ref{eq: ass U= gene bound for expo}} >0$ such that
\begin{align}
\sup_{\xi \in \Bc_{1}} \left\{\E^{\widehat\Q}[e^{\iota \int_0^T \xi^\top_t d\theta_t}] \right\}+ \E^{\widehat{\Q}}[e^{\iota \langle \theta\rangle_T}+e^{\iota\langle S \rangle_T}] \le C_{\ref{eq: ass U= gene bound for expo}},   \label{eq: ass U= gene bound for expo}
\end{align}
\end{assumption}

\begin{remark} 
\begin{itemize}
\item[(i)] The existence of the small exponential moments in \eqref{eq: ass U= gene bound for expo} implies, in particular, that there exist $C,R,\eta>0$ such that, for $\delta^3=\varepsilon \in (0,\eta)$ and all $(\xi^{j})_{j\le 3}\subset  \Bc_{1}$: 
\begin{equation}\label{eq:old}
 \E^{\widehat\Q}[  e^{ R |\xi^{1}_{T}|\bar\Rc_{\delta,\varepsilon}(\xi^{2},\xi^{3})_{T}}]<C,
 \end{equation}
 where $\bar\Rc_{\delta,\varepsilon}(\xi^{2},\xi^{3})_{T}$ is defined in \eqref{eq:Rbar}.

\item[(ii)] Suppose that the frictionless optimizer $\widehat \theta$ is of the form $d\widehat \theta_{t}=\mu^{\widehat\theta}_{t}dt+dM_{t}$ for some continuous $\widehat\Q$-local martingale $M$.
Then, the elementary estimate $\exp(|x|)\leq \exp(x)+\exp(-x)$, the Novikov-Kazamaki condition, and H\"older's inequality show that \eqref{eq: ass U= gene bound for expo} holds, in particular, if
\begin{equation}\label{eq:smallmoment}
\mathbb{E}^{\widehat\Q}\left[ e^{\kappa \int_0^T |\mu^{\widehat\theta}_t|dt}+ e^{\kappa \langle \widehat\theta \rangle_T} +e^{\kappa\langle S \rangle_T}\right] <\infty,
\end{equation}
for some arbitrarily small constant $\kappa>0$.
\end{itemize}
\end{remark}

\begin{example}\label{ex:ko96}
The bound~\eqref{eq:smallmoment} holds, e.g., for exponential utility maximization in the portfolio choice model with mean-reverting models studied by \cite{kim.omberg.96}. In (the arithmetic version) of their model,\footnote{Without transaction costs, the arithmetic and geometric versions of the model are equivalent for portfolio optimization because they span the same payoff spaces as long as the risk premium of the risky asset remains unchanged. In contrast, the precise specification matters with transaction costs. A different parametrization that covers the geometric version of the model of \cite{kim.omberg.96} is discussed in Section~\ref{sec: TC on amount}.}   the volatility $\sigma^S$ of the risky asset is constant whereas its expected returns have Ornstein-Uhlenbeck dynamics:
$$
d\mu^S_t= \lambda(\bar{\mu}^S-\mu^S_t)dt+\sigma^\mu dW^\mu_t,
$$
for constants $\lambda>0$, $\sigma^\mu \geq 0$, $\bar{\mu}^S \in \mathbb{R}$, and a $\mathbb{P}$-Brownian motion $W^\mu$ that has constant correlation $\varrho \in [-1,1]$ with the Brownian motion $W$ driving the risky returns. The optimal strategy for an exponential utility $U(x)=-e^{-rx}$ in this model is of the following form~\cite{kim.omberg.96}:
$$
\widehat\theta_t   = \frac{\mu^S_t}{r (\sigma^S)^2} +\frac{\varrho \sigma^\mu}{r\sigma^S}(B(t)+C(t)\mu^S_t),
$$
for nonpositive, smooth functions $B$, $C$ satisfying some Riccati equations. Accordingly, the quadratic variation of the frictionless optimizer $\widehat\theta$ is deterministic like for the returns process, so that these two processes evidently have finite exponential moments of all orders. To verify \eqref{eq:smallmoment}, it therefore remains to show that the drift of $\widehat\theta$ also has small exponential moments. This needs to be checked under the dual martingale measure $\widehat{\Q}$, whose density process can be derived by differentiating the value function computed in \cite{kim.omberg.96}. It follows that, under $\widehat{\Q}$, the frictionless optimizer $\widehat\theta$ is still an Ornstein-Uhlenbeck process with Gaussian distribution. Its volatility, mean-reversion level and speed are time-dependent, but bounded since they are determined by the solutions of well-behaved Riccati equations. As a result, \eqref{eq:smallmoment} is satisfied because $\widehat\theta$ is Gaussian. 
\end{example}

Under our integrability conditions, we have the following lower bound for the frictional value function:

 \begin{theorem}\label{thm: utility gene} 
 Suppose the frictionless optimal strategy $\widehat{\theta}$ is a continuous semimartingale and define the frictional tracking portfolio $\vartheta$ as in Lemma~\ref{lem: estimate L1} for $\theta=\widehat\theta$ and with $\delta^{3}:=\varepsilon\in (0,1)$. Suppose Assumptions \ref{ass: Lp integrability}, \ref{ass: Lp integrability dS/S}, \ref{ass : bounded risk aversion}, and \ref{ass: U=gene} hold for some $p>2$. Then, there exists a constant $C>0$ that does not depend on $\varepsilon \in (0,\eta)$, such that
\begin{align*}
u\big(\widehat\theta\big)-u^{\eps}(\vartheta)& \le  C\;\varepsilon^{2/3}.
\end{align*}
Moreover, $\vartheta \in \mathcal{A}^\varepsilon$, so that this estimate yields the following lower bound for the frictional value function:
\begin{equation}\label{eq:lowerbound}
v - C\;\varepsilon^{2/3} \le v^\varepsilon,
\end{equation}
for some $C>0$ that does not depend on $\varepsilon\in (0,1)$.
\end{theorem}

\proof  
Set $\Delta^{\eps}_T:=U(X^{\vartheta,\varepsilon}_T)-U(X^{\widehat\theta}_T)$.  Then, there exists  $\zeta^{\eps}_T$ that takes values between $X^{\vartheta,\varepsilon}_{T}$ and $X^\theta_{T}$ such that
\begin{align}
\E[\Delta^{\eps}_{T}]&=\mathbb{E}\left[U' (X^{\widehat\theta}_{T})\left(X^{\vartheta,\varepsilon}_{T}-X^{\widehat\theta}_{T}\right)+ \frac12 U{''}(\zeta_T^{\eps})\left(X^{\vartheta,\varepsilon}_{T}-X^{\widehat\theta}_{T}\right)^{2}\right] \notag\\
&\ge \alpha\E^{\widehat{\Q}}\left[\left(X^{\vartheta,\varepsilon}_{T}-X^{\widehat\theta}_{T}\right) + \frac12 \frac{U''(\zeta^{\eps})}{U'(X^\theta_{T})} \left(X^{\vartheta,\varepsilon}_{T}-X^{\widehat\theta}_{T}\right)^{2}\right], \label{eq:concave}
\end{align}
where we used the first-order condition \eqref{eq:FOC} and the notation $\alpha:=\E[U'(X^{\widehat\theta}_{T})]$. Now observe that Assumption \ref{ass : bounded risk aversion} implies  
$$
\left|\ln\left(\frac{U'(x)}{U'(y)}\right)\right|=\left|\int_{x}^{y} \frac{U''(z)}{U'(z)}dz\right|\le R |x-y|, \quad \mbox{for all $x,y \in \R$},
$$
so that 
$$
 \frac{U{''}(\zeta^{\eps})}{U'(X^{\widehat\theta}_{T})}=\frac{U{''}(\zeta^{\eps})}{U'(\zeta^{\eps})}\frac{U{'}(\zeta^{\eps})}{U'(X^{\widehat\theta}_{T})}\ge - R e^{R|\zeta^{\eps}- X^{\widehat\theta}_{T}|}.
$$
Together with \eqref{eq:concave}, this shows that 
\begin{align*}
\E[\Delta^{\eps}_{T}]& \ge \alpha \E^{\widehat{\Q}}\left[\left(X^{\vartheta,\varepsilon}_{T}-X^{\widehat\theta}_{T}\right)-\frac{R}{2}e^{R|\zeta^{\eps}_T-X^{\widehat\theta}_{T}|}\left(X^{\vartheta,\varepsilon}_{T}-X^{\widehat\theta}_{T}\right)^{2}\right].
\end{align*}
Now observe that $\zeta_T^{\eps}$ defined above is of the form $\lambda X^{\vartheta,\varepsilon}_{T}+(1-\lambda) X^{\widehat\theta}_{T}$ for some random variable~$\lambda$ with values in $[0,1]$. Thus $\zeta^{\eps}_T-X^{\widehat\theta}_{T} = \lambda (X^{\vartheta,\varepsilon}_{T}-X^{\widehat\theta}_{T})$, and it follows from Theorem~\ref{thm: estimate V-Veps} that 
\begin{align*}
\E[\Delta^{\eps}_{T}]& \ge \alpha\E^{\widehat\Q}\left[ \left(X^{\vartheta,\varepsilon}_{T}-X^{\widehat\theta}_{T}\right)-\frac{R}{2}e^{R\lambda  \bar \Rc_{\delta,\varepsilon}(\xi,\xi')_{T}}\left(X^{\vartheta,\varepsilon}_{T}-X^{\widehat\theta}_{T}\right)^{2}\right]
\end{align*}
for some $\xi,\xi'\in \Bc_{1}$. For $\delta=\varepsilon^{1/3}$, Remark~\ref{rem:martingale}  and Theorem \ref{thm: estimate V-Veps} together with \eqref{eq: ass U= gene bound for expo} and H\"older's inequality in turn give
\begin{align*}
\E[\Delta^{\eps}_{T}]& \ge -C\;\varepsilon^{\frac23},
\end{align*}
for some constant $C>0$ that does not depend on $\varepsilon\in (0,\eta)$. 

It remains to establish that the frictional strategy $\vartheta$ is admissible, i.e., that its wealth process $X^{\vartheta,\varepsilon}$ is a supermartingale under any absolutely continuous local martingale measure $\Q$, which has finite relative entropy with respect to the physical probability $\P$. In view of \cite[Lemma E.5]{herdegen.muhlekarbe.17}, it suffices to check that $\int_0^\cdot \vartheta_t dS_t$ is a $\Q$-supermartingale and the corresponding transaction costs $\varepsilon|\vartheta|_T$ are $\Q$-integrable. 

Since $\int_0^\cdot \widehat{\theta}_t dS_t$ is a $\Q$-supermartingale by admissibility of the frictionless optimizer $\widehat{\theta}$, it suffices to show that $\int_0^\cdot (\vartheta_t-\widehat{\theta}_t)dS_t$ is a $\Q$-martingale. Since the price process $S$ is a continuous local $\Q$-martingale, this follows if we can establish that $\E^{\Q}[\int_0^T (\vartheta_t-\widehat{\theta}_t)^2 d\langle S\rangle_t]<\infty$, cf.~\cite[Theorem~I.4.40]{js.03}. As $\vartheta-\widehat\theta$ is uniformly bounded by construction, it therefore suffices to show $\E^{\Q}\left[\langle S \rangle_T\right] < \infty$ for all equivalent martingale measures $\Q$ with finite relative entropy. In view of \cite[Lemma 3.5]{delbaen.al.02} and \cite[Theorem 2.2]{cziszar.75}, this holds if  $\E^{\widehat{\Q}}[\exp(\iota\langle S \rangle_T)]<\infty$ for some arbitrarily small $\iota$, which is part of the integrability conditions in Assumption~\ref{ass: U=gene}.

We now turn to the $\Q$-integrability of the transaction costs $\varepsilon|\vartheta|_T$. By the pathwise bound \eqref{eq: bound L in terms of processes} for $\delta=\varepsilon^{1/3}$ as well as \cite[Lemma 3.5]{delbaen.al.02} and \cite[Theorem 2.2]{cziszar.75}, this holds if 
$$
\sup_{\xi \in \mathcal{B}_1}\E^{\widehat{\Q}}\left[\exp\left(\iota \int_0^T \xi_t^{\top} d\theta_{t}\right)+\exp\left(\iota\langle \theta\rangle_T \right)\right]<\infty
$$
for some arbitrarily small $\iota>0$ as we have assumed in Assumption~\ref{ass: U=gene}. This shows that $\vartheta$ is indeed admissible, completing the proof.
\ep

\section{It\^{o} Decomposition of the Frictionless Optimizer Using Malliavin Calculus}\label{s:malliavin}
 
 In this section, we explain how to verify the conditions imposed on the frictionless optimizer in Theorem~\ref{thm: utility gene}. More specifically, we show that a simple application of the Clark-Ocone formula provides sufficient conditions in terms of the primitives of the model, avoiding the need for abstract assumptions on the frictionless optimizer.
 
To ease notation, we focus on a simple one-dimensional, time-homogeneous Markov model where the asset price process $S$ is the solution of a stochastic differential equation,     
 \begin{align}\label{eq: def X} 
 S=S_0+\int_{0}^{\cdot} \mu(S_{t})dt+\int_{0}^{\cdot} \sigma(S_{t})dW_{t}.
 \end{align}
Here, $S_{0}\in \R$, $W$ is a one-dimensional standard Brownian and $\mu$, $\sigma$ are  globally Lipschitz maps taking values in $\R$ and $(0,\infty)$, respectively. 

\begin{remark}
Adding a time dependency in the coefficients or considering a multi-dimensional setting would not change the nature of the analysis. In principle, our approach could also be extended to non-Markovian settings, but we do not pursue this here since the corresponding assumptions for Malliavin differentiability would be rather involved and abstract. 
\end{remark}

For our analysis based on Malliavin calculus, the primitives of the model need to be sufficiently regular. The following conditions are sufficient; for clarity, we do not strive for minimal assumptions. 

 \begin{assumption}\label{ass: for the control of theta} The maps $\lambda:=\sigma^{-1}\mu$, $\sigma$ and $\sigma^{-1}$ are twice continuously differentiable with bounded derivatives of order $0, 1, 2$. Moreover, the utility function $U$ satisfies Assumption \ref{ass : bounded risk aversion}, $U\in C^{3}(\R)$, and $U'''/U''$ is bounded.
 \end{assumption}

For a bounded market-price of risk $\lambda$ as in Assumption~\ref{ass: for the control of theta}, the measure $\widehat{\Q} \sim \P$ with density
 \begin{align}\label{eq: def N}
 \frac{d\widehat\Q}{d\P}=e^{N}, \quad \mbox{ where } N:=\frac12\int_{0}^{T} \lambda(S_{t})^{2}dt+\int_{0}^{T} \lambda(S_{t})dW^{\hat \Q}_{t},
 \end{align}
 is the unique martingale measure for $S$, and $W^{\widehat\Q}:=W-\int_{0}^{\cdot} \lambda(S_{t})dt$ is a $\widehat\Q$-Brownian motion. As $\widehat\Q$ trivially is the minimizer of the dual problem, it is linked to the optimal strategy $\widehat{\theta}$ for the primal problem by the first-order condition \eqref{eq:FOC}. Since the derivatives of the utility function $U$ and its convex conjugate $\tilde U(y):=\sup_{x \in \mathbb{R}}\{U(x)-xy\}$, $y\in \R$, are related by $\tilde{U}'(x)=-(U')^{-1}(x)$, this means that we can  find $h>0$ such that 
$$
X^{\widehat\theta}_{T}=-\tilde U'(H ), \quad \mbox{where } H:=h e^{N}.
$$ 
Assumption~\ref{ass: for the control of theta} in turn also guarantees the integrability of the optimal frictionless wealth process:

\begin{lemma}\label{lem:bound}
Under Assumption~\ref{ass: for the control of theta}, we have $\tilde U'(H )\in \Lb_{2}(\widehat\Q)$.
\end{lemma}

\proof 
As $\tilde{U}'(x)=-(U')^{-1}(x)$, observe that the derivative of $x\mapsto \tilde{U}'(\exp(x))$ is given by $x\mapsto \exp(x)/U''[(U')^{-1}(\exp(x))]$. Since $\exp(x)=U'[(U')^{-1}(\exp(x))]$, the derivative of $x\mapsto \tilde{U}'(\exp(x))$ is therefore bounded because $U'/U''$ is bounded by assumption. Thus, $x \mapsto \tilde{U}'(\exp(x))$ has at most linear growth and the integrability of $\tilde{U}'(H)$ in turn follows from Assumption~\ref{ass: for the control of theta}.
\ep\\

We can now establish the main result of this section, which shows that under Assumption~\ref{ass: for the control of theta}, the frictionless optimizer is not only a continuous semimartingale but in fact an It\^o process with bounded drift and diffusion coefficients. In particular, Theorem~\ref{thm: utility gene} is applicable in this case.

 \begin{proposition}\label{prop: theta bounded} Let Assumption \ref{ass: for the control of theta} hold. Then, the frictionless optimizer $\widehat{\theta}$ is bounded and of the form 
 $$
\theta =\widehat{\theta}_{0}+\int_{0}^{\cdot} \alpha_{t}dt +\int_{0}^{\cdot} \gamma_{t} dW^{\widehat \Q}_{t}\bru{,}
$$ 
where $\widehat{\theta}_{0}\in \R$ and $\alpha$, $\gamma$ are bounded, adapted processes. 
 \end{proposition}
 
\proof \emph{Step 1}: we first prove that $\widehat{\theta}$ is  bounded by applying the Clark-Ocone formula. We denote by $D_{t}$ the time-$t$ Malliavin derivative operator with respect to $W^{\widehat\Q}$. It follows from \reff{eq: def X}, \reff{eq: def N} and    \cite[Theorem 2.2 and p.104]{nualart} (applied to the two-dimensional diffusion process $(S,N)$) that 
 $$
D_{t}N=\int_{t}^{T} (\lambda\lambda')(S_{s})D_{t}S_{s}ds+\int_{t}^{T} \lambda'(S_{s})D_{t}S_{s}dW^{\widehat\Q}_{s}+ \lambda(S_{t}),
$$ 
where  
 \begin{align}\label{eq: Dt X}
  D_{t}S=\sigma(S_{t})e^{-\frac12 \int_{t}^{\cdot} |\sigma'(S_{s})|^{2} ds+\int_{t}^{\cdot} \sigma'(S_{s}) dW^{\widehat\Q}_{s}}.
 \end{align}
Hence, 
\begin{align}\label{eq: DtN}
D_{t}N= \sigma(S_{t})\Ec_{t}^{-1}\left[\bar N_{T}-\bar N_{t}\right]+ \lambda(S_{t}), 
\end{align}
with  
\begin{align*}
\bar N:=\int_{0}^{\cdot} (\lambda\lambda')(S_{s})\Ec_{s}ds+\int_{0}^{\cdot} \lambda'(S_{s})\Ec_{s}dW^{\widehat\Q}_{s}\quad \mbox{ and }\quad
\Ec:=e^{-\frac12\int_{0}^{\cdot} |\sigma'(S_{s})|^{2}ds+\int_{0}^{\cdot} \sigma'(S_{s})dW^{\widehat\Q}_{s}}.
\end{align*}
Note that, using standard estimates, our bounds on $\sigma, \sigma'$, $\lambda$ and $\lambda'$ imply that, for all $p\ge 1$, 
\begin{align}
&\sup_{t\in [0,T]}\|\E^{\widehat\Q}\Big[\sup_{s \in [t,T]}|\Ec_s/\Ec_{t}|^{p}|\Fc_{t}\Big]\|_{\Lb_{\infty}}<\infty,\label{eq: bound xis/xit}\\
&\sup_{t\in [0,T]}\|\E^{\widehat\Q}[| (\bar N_{T}-\bar N_{t})/\Ec_{t}|^{p}|\Fc_{t}]\|_{\Lb_{\infty}}<\infty,\label{eq: bound ecart bar N/xit}\\
&\sup_{t\in [0,T]}\|\E^{\widehat\Q}[| D_{t}N|^{p}|\Fc_{t}]\|_{\Lb_{\infty}}<\infty\label{eq: bound Dt N},\\
&\E^{\widehat\Q}[H^{p}]<\infty.\label{eq: bound H}
\end{align}
Moreover, $X^{\widehat\theta}_{T}=-\tilde U'(H)$ and the chain-rule formula \cite[Proposition 1.2.3, Lemma 1.2.3]{nualart} imply  
$$
D_{t}X^{\widehat\theta}_{T}=-H\tilde U''(H)D_{t}N=F\Big(X^{\widehat\theta}_{T}\Big)D_{t}N,
$$ 
where $F:=U'/U''$. Recall that $F$ is bounded by assumption and $\tilde U'(H )\in \Lb_{2}(\widehat\Q)$ by Lemma~\ref{lem:bound}. In view of  \reff{eq: bound Dt N}, it follows that  $X^{\widehat\theta}_{T}$ belongs to the Malliavian space ${\mathbb D}^{1,2}$, see \cite[p.27]{nualart}, and that 
\begin{align}\label{eq: bound term 2}
 \sup_{t\le T}\|\E^{\widehat\Q}\Big[|D_{t}X^{\widehat\theta}_{T}|^{2}|\Fc_{t}\Big]\|_{\Lb_{\infty}}=\sup_{t\le T}\|\E^{\widehat\Q}\Big[|F\big(X^{\widehat\theta}_{T}\big)D_{t}N|^{2}|\Fc_{t}\Big]\|_{\Lb_{\infty}}<\infty.
\end{align}
One can then apply the Clark-Ocone formula \cite[Proposition 1.3.14]{nualart}  to obtain
\begin{align*}
\widehat{\theta}_t \sigma(S_t)=\E^{\widehat\Q}\Big[D_{t}X^{\widehat\theta}_{T}|\Fc_t\Big], \quad t \in [0,T].
\end{align*}
Since $\sigma^{-1}$ is bounded, \eqref{eq: bound term 2} in turn shows that $\widehat\theta$ is indeed bounded.

\emph{Step 2}: next, we prove that $\widehat\theta$ has a bounded quadratic variation. Set 
$$\tilde F:=F((U')^{-1})$$
and recall that $U'(X^{\widehat\theta}_{T})=H$. Then, it follows from Step 1 that 
 \begin{align*}
 \widehat\theta_{t}=\E^{\widehat\Q}\Big[D_{t}X^{\widehat\theta}_{T}|\Fc_{t}\Big] \sigma(S_{t})^{-1}
&= \Ec_{t}^{-1}(A_{t}-B_{t}\bar N_{t})+ (\sigma^{-1} \lambda)(S_{t})B_{t},\quad t\in [0,T],
\end{align*}
where 
\begin{align*}
A_{t}&:=\E^{\widehat\Q}[\tilde F( H)\bar N_{T}|\Fc_{t}]=\E^{\widehat\Q}[\tilde F( H)\bar N_{T}]+\int_{0}^{t} \phi^{A}_{s}dW^{\widehat\Q}_{s},\\
B_{t}&:=\E^{\widehat\Q}[\tilde F( H)|\Fc_{t}]=\E^{\widehat\Q}[\tilde F( H)]+\int_{0}^{t}\phi^{B}_{s}dW^{\widehat\Q}_{s}. 
\end{align*}
Here, $(\phi^{A},\phi^{B})$ are obtained from the Clark-Ocone formula \cite[Proposition 1.3.14]{nualart}  and the chain-rule formula \cite[Proposition 1.2.3, Lemma 1.2.3]{nualart}  : 
\begin{align*}
\phi^{A}_{s}:=\E^{\widehat\Q}[\tilde F'( H) H D_{s}N\bar N_{T}+\tilde F( H)D_{s}\bar N_{T}|\Fc_{s}]
\;,\;\phi^{B}_{s}:=\E^{\widehat\Q}[\tilde F'( H) H D_{s}N|\Fc_{s}], \quad s\in [0,T].
\end{align*}
Again, the required integrability conditions can be easily deduced from Assumption \ref{ass: for the control of theta} by arguing as in Step 1. As a consequence, 
\begin{align*}
\frac{d}{dt}\langle \widehat\theta - (\sigma^{-1} \lambda)(S)B\rangle_{t} =& |\Ec_{t}^{-1}(\phi^{A}_{t}-\phi^{B}_{t}\bar N_{t})-B_{t}\lambda'(S_{t})-(A_{t}-B_{t}\bar N_{t})\Ec_{t}^{-1}\sigma'(S_{t})|^{2}
\end{align*}
where, by \reff{eq: DtN},  
\begin{align*}
\Ec_{t}^{-1}(\phi^{A}_{t}-\phi^{B}_{t}\bar N_{t}):=& \sigma(S_{t})\Ec_{t}^{-2}\E^{\widehat\Q}[\tilde F'( H) H (\bar N_{T}-\bar N_{t})^{2}|\Fc_{t}]+\Ec_{t}^{-1}\E^{\widehat\Q}[\tilde F( H)D_{t}\bar N_{T}|\Fc_{t}]\\
&+\lambda(S_{t})\Ec_{t}^{-1}\E^{\widehat\Q}[\tilde F'( H) H (\bar N_{T}-\bar N_{t})|\Fc_{t}]
\end{align*}
and 
$$
(A_{t}-B_{t}\bar N_{t})\Ec_{t}^{-1}=\E^{\widehat\Q}[\tilde F( H)(\bar N_{T}-\bar N_{t})|\Fc_{t}]\Ec_{t}^{-1}.
$$
The identity 
\begin{align*}
|\tilde F'( H) H|&=\left|\frac{U'}{U''}\Big(X^{\widehat\theta}_{T}\Big)\left(1-\frac{U'U'''}{(U'')^{2}}\Big(X^{\widehat\theta}_{T}\Big)\right)\right|
\end{align*}
combined with our assumptions ensures that $\tilde F'( H) H$ is bounded. This is also the case for $\Ec_{t}^{-2}\E^{\widehat\Q}[(\bar N_{T}-\bar N_{t})^{2}|\Fc_{t}]$ by \reff{eq: bound ecart bar N/xit}, and for $\tilde F(H)$ by assumption. Moreover,  
\begin{align*}
\Ec_{t}^{-1}D_{t}\bar N_{T}=\lambda'(S_{t})+\int_{t}^{T} \Ec_{t}^{-1}D_{t}[(\lambda\lambda')(S_{s})\Ec_{s}]ds+\int_{t}^{T} \Ec_{t}^{-1}D_{t}[\lambda'(S_{s})\Ec_{s}]dW^{\widehat\Q}_{s} 
\end{align*}
where
$$
D_{t}\Ec_{s}=\Ec_{s} \times \left(\sigma'(S_{t})-\int_{t}^{s} (\sigma'\sigma'')(S_{s})D_{t}S_{s} ds +\int_{t}^{s} \sigma''(S_{s})D_{t}S_{s} dW^{\widehat\Q}_{s}\right), 
$$
so that \reff{eq: Dt X}, \reff{eq: bound xis/xit},  Assumption \ref{ass: for the control of theta}, and the chain-rule formula imply that 
$$
\sup_{t\in [0,T]}\|\E^{\Q}[|\Ec_{t}^{-1}D_{t}\bar N_{T}||\Fc_{t}]\|_{\infty}<\infty. 
$$
The above bounds combined with  Assumption \ref{ass: for the control of theta} yield that $ \langle \widehat\theta - (\sigma^{-1} \lambda)(S)B\rangle$ and  $ \langle   (\sigma^{-1} \lambda)(S)B\rangle$ are bounded.  By polarization, it follows that $ \langle \widehat\theta\rangle $ is bounded as well. 

\emph{Step 3}: it remains to prove that the drift part of $\widehat\theta$ has a bounded density. Recall from Step~2 that the frictionless optimizer can be represented as 
\begin{align*}
 \widehat\theta_{t}= \Ec_{t}^{-1}(A_{t}-B_{t}\bar N_{t})+ [\sigma^{-1} \lambda](S_{t})B_{t}.
\end{align*}
After applying It\^{o}'s formula, our assumptions and similar arguments as in Step 2 show that its dynamics are of the following form:
$$
\widehat\theta_t =\widehat\theta_{0}+\int_{0}^{t} \beta_{s}ds +\int_{0}^{t} \gamma_{s} dW^{\widehat\Q}_{s}, \quad t \in [0,T],
$$ 
with $\widehat\theta_{0}\in \R$, $|\gamma|^{2}=d\langle \widehat\theta\rangle/dt$ and 
\begin{align*}
\beta_{t}:=&(A_{t}-B_{t}\bar N_{t})\Ec_{t}^{-1}|\sigma'(S_{t})|^{2}-\Ec_{t}^{-1}\sigma'(S_{t})(\phi^{A}_{t}-\phi^{B}_{t}\bar N_{t})\\
&- B_{t}[(\lambda\lambda')(S_{t})-(\lambda'\sigma')(S_{t})+\lambda'(S_{t}) \phi^{B}_{t}]\\
&+\Lc (\sigma^{-1} \lambda)(S_{t})B_{t}+[(\sigma^{-1} \lambda)'\sigma](S_{t})\phi^{B}_{t}.
\end{align*}
Here, $\Lc$ denotes the Dynkin operator associated to $S$ under $\widehat\Q$. Similar arguments as in Step~2 -- now also using the boundedness of the second-order derivatives -- in turn show that $\beta$ is indeed bounded. This completes the proof.
\ep

\section{Strategies Parametrized in Monetary Amounts}\label{sec: TC on amount}
  
For simplicity, our results are presented in the case where the controls $\theta$ and $\vartheta$ describe the numbers of shares of the risky assets held in the portfolio, and transaction costs are levied on the number of shares transacted. A similar analysis can also be conducted when the controls represent monetary amounts invested and transaction costs are proportional to dollar amounts traded. We now outline how to adapt the arguments from Sections~\ref{s:bounds} and \ref{s:utility} in this case.

In order to parametrize strategies in terms of monetary risky positions, suppose that the price process $S$ is a $(0,\infty)^{d}$-valued continuous semimartingale. The frictionless target strategy $\theta$ is in turn assumed to be a continuous semimartingale such that $\theta/S$ is $S$-integrable.\footnote{Here, we use the notation $x/y=(x^{i}/y^{i})_{i\le d}$.} Here, $\theta^{i}$ now represents the amount of money invested in the risky asset $i$, so that the corresponding frictionless wealth process is
$$
X^\theta_t :=X_{0}+\int_{0}^t  (\theta_{s}/S_{s})^{\top} dS_{s}, \quad t \in [0,T].
$$ 
Now suppose trades incur costs proportional to the monetary amount transacted. Then, the frictional wealth process is 
$$
X^{\vartheta,\eps}_t :=X_{0}+\int_{0}^{t} { (Y^{\vartheta}_{s}/S_{s})}^{\top}dS_{s}- \eps\int_{0}^{t}d|\vartheta|_{s}- \1_{\{T\}}\eps {|Y^\vartheta_{T}|},
$$
where the amounts of money invested into the risky assets follows
$$
Y^\vartheta_{t}:=\int_{0}^{t}{ (Y^{\vartheta}_{s}/S_{s})}^{\top}dS_{s}+\vartheta_{t}.
$$
These positions are controlled through the continuous bounded-variation process $\vartheta^{i}$, which describes the cumulative amount of money transferred to the position in the corresponding asset so far. In this setting, the Skorokhod problem studied in Lemma \ref{lem: estimate L1} becomes 
\begin{equation*}
\left\{\begin{array}{c}
\theta-{Y^{\vartheta}}\in [-\delta,\delta]^{d} \mbox{ on } [0,T],\\
\sum_{i=1}^{d}\left(\int_{0}^{T}\1_{\{\theta^{i}_{t}-{Y}^{\vartheta,i}_{t}=\delta \}}d\vartheta^{i +}_{t}+\int_{0}^{T}\1_{\{\theta^{i}_{t}-{Y}^{\vartheta,i}_{t}=-\delta\} }d\vartheta^{i-}_{t}\right)=0.
\end{array}\right.
\end{equation*} 
The arguments used to prove Lemma~\ref{lem: estimate L1} can in turn be adapted to control $|\vartheta|$ also in this setting. Indeed, define $\vp$ as in the proof of Lemma \ref{lem: estimate L1} and set $Z:=(\theta -Y^{\vartheta})/\delta$. Then, focusing on a single risky asset ($d=1$) without loss of generality, it follows from It\^{o}'s lemma and the identities $-\vp'(-1)=\vp'(1)=1$ that 
 \begin{align*}
 \vp(Z_{t})=&\vp(Z_{0})+\frac1\delta\left(\int_{0}^{t}   \vp'( Z_{s})d(\theta-Y^{\vartheta})_{s}
+\frac1{2\delta } \int_{0}^{t}  \vp{''}(Z_{s})d\langle \theta-Y^{\vartheta}\rangle_{s}\right)\nonumber\\
=&\vp(Z_{0})+\frac1\delta\left(\int_{0}^{t}   \vp'( Z_{s})d\theta_{s}-\int_{0}^{t}   \vp'( Z_{s})(Y^{\vartheta}_{s}/S_{s})dS_{s}-|\vartheta|_{t}\right)\\
&+\frac1{2\delta^{2}}\left( \int_{0}^{t}  \vp{''}(Z_{s})d\langle \theta\rangle_{s}+\int_{0}^{t}  \vp{''}(Z_{s})(Y^{\vartheta}_{s}/S_{s})^{2}d\langle S\rangle_{s}-2\int_{0}^{t}  \vp{''}(Z_{s})(Y^{\vartheta}_{s}/S_{s})d\langle S,\theta\rangle_{s})\right).
 \end{align*}
 One can then use the bounds on $\vp$ and its derivatives ($|\vp|\vee |\vp'|\vee |\vp{''}|\le 1$ on $[-1,1]$) as well as $(Y^{\vartheta}-\theta)/\delta\in \Bc_{1}$ to obtain the following estimate:
 \begin{align*}
 |\vartheta|_{t}\le & \;\delta\left(2+\int_{0}^{t}\xi^{1}_{s} dM_{s}+\frac12\int_{0}^{t}  d\langle M\rangle_{s} \right)\\
 &+ \int_{0}^{t}   \xi^{2}_{s}d\theta_{s}+\int_{0}^{t}\xi^{3}_{s}\theta_{s}dM_{s}+ \int_{0}^{t}\xi^{4}_{s}\theta_{s} {d\langle M\rangle_{s}} +\int_{0}^{t}   \xi^{5}_{s}{d\langle  M,\theta\rangle_{s}} \\
 &+\frac1\delta\left(\frac12 \langle \theta\rangle_{t} +\frac12  \int_{0}^{t}\theta_{s}^{2} {d\langle M\rangle_{s}}  +\int_{0}^{t}\xi^{6}_{s}\theta_{s}  {d\langle  M,\theta\rangle_{s}}  \right).
 \end{align*} 
Here, $\xi^{1},\ldots,\xi^{6}\in \Bc_{1}$, and $M$ denotes the returns process with dynamics $dM_t=dS_t/S_t$.

 The counterparts of Corollary \ref{cor: estimate L} and Theorem \ref{thm: estimate V-Veps} in turn follow under integrability conditions similar but somewhat more involved than Assumptions \ref{ass: Lp integrability} and \ref{ass: Lp integrability dS/S}. An analogue of Theorem \ref{thm: utility gene} can also be obtained under conditions similar to Assumption \ref{ass: U=gene}.
 
 In particular, arbitrarily small exponential moments of the primitives of the model are still sufficient to derive the lower bound \eqref{eq:lowerbound}. In particular, this allows to cover the geometric version of the model of~\cite{kim.omberg.96} where the frictionless target strategy $\theta=\widehat{\theta}$ is Gaussian, compare Example~\ref{ex:ko96}.  In contrast, the integrability conditions imposed in previous papers are only satisfied in this context if the time horizon $T$ is sufficiently small. Indeed, \cite[Condition (3.2)]{kallsen.li.13} or \cite[Condition (A.2)]{herdegen.muhlekarbe.17} require the existence of a specific exponential moment for $\int_0^T \widehat\theta_t dM_t$ or $\int_0^T \widehat{\theta}^2_t dt$, which both involve a squared Ornstein-Uhlenbeck process for the model of \cite{kim.omberg.96}. Therefore, these conditions only hold if the time horizon $T$ is not too large. 

\bibliographystyle{plain}

\end{document}